\documentclass[letterpaper,twocolumn,10pt]{article}

\usepackage{usenix2019_v3}

\usepackage[T1]{fontenc}
\usepackage[utf8]{inputenc}

\usepackage{amsmath}
\usepackage{amssymb}
\usepackage{amsfonts}
\usepackage{amsthm}

\usepackage{graphicx}
\usepackage{booktabs}
\usepackage{array}
\usepackage{tabularx}
\usepackage{subcaption}

\usepackage{tikz}
\usepackage{algorithm}
\usepackage[noend]{algpseudocode}

\usepackage{xcolor}
\usepackage{xspace}
\usepackage{microtype}
\usepackage{listings}

\usepackage{xcolor}
\usepackage{xspace}
\usepackage{microtype}
\newcommand{\sys}{CoLoc\xspace}
 \newcommand{\figspace}{\vspace{-2mm}}

\begin{document}


\date{}

\title{\Large \bf
Cross-Modal Data Co-Locality for Heterogeneous Storage Systems
}

\author{
Abdullah Al-Mamun\\
Augusta University\\
aalmamun@augusta.edu
}

\maketitle


\begin{abstract}
Distributed vector stores increasingly serve multimodal workloads whose
representations are jointly accessed but independently partitioned and
placed. As a result, cross-modal physical locality remains largely
unoptimized. Multimodal queries can incur remote accesses and cross-tier
data movement even when each modality's ANN search remains efficient.
We present \sys{}, a storage-management layer that makes cross-modal
access locality a first-class data-layout signal. A Cross-Modal Access
Graph combines semantic affinity, co-access, and temporal behavior. A
movement-aware optimizer places frequently co-accessed vectors under
capacity and topology constraints. A migration manager adapts placement
as workloads shift by weighing locality gains against migration cost. A
confidence-gated prefetcher hides remaining slow-tier latency. \sys{}
operates above existing ANN indexes without modifying embeddings, search,
or retrieval semantics.
We deploy \sys{} on a multi-tier AWS testbed spanning GPU, NUMA-DRAM,
NVMe, and S3. \sys{} substantially reduces data movement and query
latency while leaving ANN retrieval unchanged. Under a scheduled workload
shift, its migration manager matches the locality of periodic global
reorganization while moving 5.9$\times$ less data and completing
3.9$\times$ faster.

\end{abstract}

\section{Introduction}
\label{sec:introduction}\figspace

Multimodal applications increasingly represent the same logical object using
multiple modality-specific embeddings, such as text, images, clinical records,
and sensor data~\cite{zhang2025gme,cui2025dekr,caffagni2025ret,duan2025fume}.
Medical workloads, for example, increasingly combine electronic health
records (EHRs), medical images, laboratory measurements, and other clinical
modalities~\cite{elsharief2025medmod}, motivating new systems support for
multimodal data storage and retrieval~\cite{lu2025mmstorage}. These
representations are often independently embedded and indexed, allowing each
modality to use an appropriate ANN structure but also allowing related
representations to become physically separated.

This separation creates a physical data-layout problem that is largely
orthogonal to ANN search efficiency. Consider a multimodal query that first
searches text or EHR vectors and subsequently accesses the corresponding
image representations. The two modality-specific ANN searches can each be
efficient, yet their results may reside on different machines, NUMA domains,
memory tiers, or storage devices. The system must then move data across the
network, NUMA hierarchy, CPU--GPU boundary, or storage hierarchy to assemble
the final result. Thus, a vector system can optimize its individual ANN
searches while still paying a substantial physical data-movement cost at the
\emph{multimodal query level}. The fundamental problem is therefore not only
where vectors should be searched, but also \emph{where heterogeneous
representations required by the same query should physically meet}.

\begin{table*}[!t]
\centering
\caption{Positioning CoLoc against representative locality-aware vector,
memory, and storage systems. }
\figspace
\label{tab:positioning}
\footnotesize
\setlength{\tabcolsep}{4pt}
\renewcommand{\arraystretch}{0.95}
\begin{tabular}{lccccc}
\hline
\textbf{System}
& \textbf{Placement Scope}
& \textbf{Locality Signal}
& \textbf{Cross-Modal}
& \textbf{Movement-Aware}
& \textbf{Adaptive Migration} \\
\hline

CoTra~\cite{zhi2025cotra}
& Vector partitions
& Search / vector similarity
& -- & -- & -- \\

RED-ANNS~\cite{chen2025redanns}
& ANN graph partitions
& Graph / vector similarity
& -- & -- & -- \\

d-HNSW~\cite{fang2026dhnsw}
& ANN data
& ANN search locality
& -- & -- & -- \\

Turbocharging~\cite{shim2025turbocharging}
& ANN index layout
& Access locality
& -- & -- & -- \\

M2T~\cite{zhong2024managing}
& Memory pages
& Memory-tier access
& -- & \checkmark & \checkmark \\

PolyStore~\cite{ren2025polystore}
& Storage objects
& Storage workload
& -- & \checkmark & -- \\

LACX~\cite{jeong2025lacx}
& Memory pages
& Shared-page access
& -- & \checkmark & \checkmark \\

\textbf{CoLoc}
& \textbf{Cross-modal placement units}
& \textbf{Co-access + semantic + temporal}
& \textbf{\checkmark}
& \textbf{\checkmark}
& \textbf{\checkmark} \\

\hline
\end{tabular} \figspace \figspace
\end{table*}

Recent vector systems increasingly optimize physical locality, but at
different levels. Distributed ANN systems such as CoTra and RED-ANNS optimize
partitioning, communication, and distributed execution
\cite{zhi2025cotra,chen2025redanns}, while d-HNSW explores vector search over
disaggregated memory~\cite{fang2026dhnsw}. Storage-aware vector systems exploit
locality within vector indexes to reduce I/O~\cite{shim2025turbocharging}, and
heterogeneous-memory systems study locality-aware placement and migration
across memory tiers~\cite{zhong2024managing,tang2024tiresias,jeong2025lacx}.
Meanwhile, vector database architectures are increasingly spanning
heterogeneous memory and storage resources~\cite{pan2024vectordb,
song2026vectorsearch,suri2026vextra}. These systems address important forms
of index, communication, or resource locality, but do not make
\emph{cross-modal co-consumption} a first-class physical placement objective.
Table~\ref{tab:positioning} summarizes the fundamental limitations of current representative locality-aware systems.

In summary,
multimodal queries induce a distinct form of physical locality:
representations that are repeatedly consumed together should preferably be
physically close, even when they belong to different indexes and modalities.
Semantic similarity alone is insufficient because semantically related
representations need not be accessed together, while observed co-access can
be sparse for emerging workloads. The placement problem must therefore
combine workload co-access with semantic and temporal evidence while
accounting for heterogeneous hardware costs.

We present \sys{}, a storage-management layer that introduces
\emph{cross-modal co-locality} as a first-class placement objective without
modifying embeddings, ANN indexes, distance metrics, or retrieval semantics.
The \emph{Cross-Modal Access Graph (CAG)} captures relationships among
independently indexed representations using cross-modal co-access, semantic
affinity, and temporal proximity. The \emph{Co-Locality Placement Optimizer
(CPO)} translates these relationships into capacity-, load-, and
topology-aware placement decisions while accounting for physical movement
cost. A topology-aware layer maps logical decisions onto GPU memory, NUMA
DRAM, NVMe, and remote storage, while the \emph{Locality-Preserving Migration
Manager (LPMM)} adapts placement under workload drift only when the expected
future locality benefit justifies migration cost.
The central distinction is simple: \emph{existing vector systems optimize
where vectors are searched; \sys{} optimizes where the heterogeneous
representations required by a query physically meet.}

In summary, the paper makes the following contributions:

\begin{itemize}
    \item \textbf{Cross-modal co-locality abstraction.}
    We introduce cross-modal co-locality as a physical data-layout primitive
    for independently indexed multimodal representations.

    \item \textbf{Movement-aware placement.}
    We develop CAG and CPO to translate cross-modal co-access, semantic and
    temporal evidence into topology- and capacity-aware placement decisions
    that minimize physical data movement.

    \item \textbf{Adaptive migration.}
    We develop LPMM to adapt placement under workload changes while
    explicitly trading future movement savings against migration cost.

    \item \textbf{System evaluation.}
    We implement \sys{} above existing ANN indexes and evaluate it on a
    real multi-tier AWS deployment spanning GPU, NUMA-DRAM, NVMe, and S3.
    \sys{} eliminates cross-tier movement entirely (73\%$\to$0\% of
    multimodal queries) and reduces end-to-end latency by up to 25\% over
    independent placement at high multimodal query intensity. LPMM adapts
    to workload shift while moving 5.9$\times$ less data and completing
    3.9$\times$ faster than periodic global reorganization; a CAG-driven
    prefetcher further reduces latency by 8.5\% by overlapping remote
    fetches with query processing. Retrieval quality (Recall@$k$) remains
    completely unaffected by placement throughout.
\end{itemize}




\section{Related Work}
\label{sec:related}\figspace 
\noindent\textbf{Newer ANN serving and multimodal query systems.} Recent
large-scale and GPU-accelerated ANN systems -- DistVS~\cite{yin2026distvs},
GIDS~\cite{gids2025}, and VecFlow~\cite{vecflow2025} -- continue the
search-execution focus already discussed in \S\ref{sec:introduction};
none reasons about placement across independently-managed indexes.
Closer in spirit is ARCADE~\cite{yang2025arcade}, a real-time engine for
hybrid, continuous queries across diverse data modalities. ARCADE
addresses \emph{query execution} across modalities within one system;
it does not address the physical co-location of independently-indexed
representations across heterogeneous storage tiers, which is where
\sys{} operates -- the two are complementary rather than competing. 

\noindent\textbf{General cost-aware data placement.} Beyond the heterogeneous
storage/memory placement model of \S\ref{sec:cpo}, similar
movement-vs-reorganization trade-offs appear in learned directory
placement~\cite{oldp2024}, Markov-clustering-based locality for
MapReduce~\cite{idaps2024}, and multicloud placement~\cite{adapt2025multicloud}.
This work is domain-agnostic; \sys{} specializes the same trade-off to
independently-indexed multimodal vectors, where the placement unit and
migration trigger are shaped by cross-modal query structure
(\S\ref{sec:query_locality}) rather than a generic access trace.
Industry movement toward object-storage-integrated vector
search~\cite{bader2025cephs3vectors} and filtered ANN search at
scale~\cite{pinecone2025filtering} reflects the same trend motivating
\sys{}: vector workloads increasingly span storage tiers by default. 

\noindent\textbf{Multimodal retrieval quality and clinical fusion.} A
separate line of work improves \emph{what} gets embedded and how
modalities are combined for retrieval quality: generative multimodal
search~\cite{kim2025genius} and noisy-correspondence-robust cross-modal
alignment~\cite{wang2026mgcs,wang2026dns,recon2025,noisycorrespondence2026}.
In the clinical domain, DrFuse~\cite{yao2024drfuse} and missing-modality
EHR embedding~\cite{pmlr-v219-lee23a} extend the fusion challenge
\S\ref{sec:introduction} raised via MedMod. All of this determines
embedding and retrieval semantics, which \sys{} explicitly does not
touch (\S\ref{sec:design}); better embeddings increase the value of
correct placement without changing where independently-indexed vectors
physically end up relative to each other -- the gap \sys{} fills. 

\noindent\textbf{VDBMS engineering maturity.} Work characterizing reliability
gaps~\cite{wang2025reliablevdbms}, real production bugs~\cite{xie2025vdbbugs},
and fundamental system concepts~\cite{taipalus2023vectordb} in vector
database management reflects growing engineering rigor in this space.
Physical placement across heterogeneous tiers remains comparatively
under-examined in that line of work; \sys{} treats it as a first-class,
independently evaluable concern (\S\ref{sec:evaluation}). \figspace \figspace

\section{Cross-Modal Co-Locality Design}
\label{sec:design}\figspace

\sys{} is a storage-management layer above existing modality-specific
embedding and ANN indexes. It does not modify embeddings, ANN search
algorithms, distance functions, index structures, or retrieval semantics.
Instead, it adapts the physical placement of heterogeneous
representations according to their joint consumption by multimodal
queries.

Cross-modal accesses are asynchronously collected by the
\emph{Cross-Modal Access Graph (CAG)}, which combines observed co-access
with semantic and temporal evidence through incremental updates. The
\emph{Co-Locality Placement Optimizer (CPO)} uses this signal to determine
movement-aware placement through an iterative group-partitioning
heuristic and maps the resulting groups onto topology- and tier-aware
resources. The \emph{Locality-Preserving Migration Manager (LPMM)}
adapts placement when workload changes justify migration. A CAG-driven
prefetcher (\S\ref{sec:prefetch}) complements these mechanisms by warming
a bounded cache ahead of predicted accesses without changing placement.
Together, these components form a closed control loop with no synchronous
coordination on the query critical path.\figspace

\subsection{Query-Induced Cross-Modal Locality}
\label{sec:query_locality}

Existing vector systems~\cite{zhi2025cotra,chen2025redanns,fang2026dhnsw,
pan2024vectordb,song2026vectorsearch,suri2026vextra,shim2025turbocharging}
reason about locality at the level of individual vectors, ANN
neighborhoods, or storage blocks -- not multimodal queries whose results
span several independently managed indexes. \sys{} instead defines
locality at the level of \emph{joint query consumption}.

Let $\mathcal{V}=\{v_1,\ldots,v_N\}$ be modality-specific
representations and $\mathcal{H}=\{h_1,\ldots,h_M\}$ available
heterogeneous resources (GPU, NUMA, NVMe, remote). A \emph{placement
unit} is the smallest independently movable storage extent exposed by
the underlying layer. Let $p(v)$ denote the resource holding $v$'s
placement unit. A multimodal query induces
$Q_i=\{v_{i1},\ldots,v_{ik_i}\}$, possibly spanning multiple modalities
and indexes. The physical separation cost of a query is
\begin{equation}
\footnotesize
C_{\mathrm{sep}}(Q_i,p) =
\sum_{\substack{u,v\in Q_i\\u<v}} D\bigl(p(u),p(v)\bigr),
\label{eq:separation}
\end{equation}
where $D(h_a,h_b)$ is the physical cost (seconds) of moving data between
$h_a,h_b$ (\S\ref{sec:topology}). Over a workload $\mathcal{Q}$ with
frequencies $\rho_i$, the expected movement cost is
\begin{equation}
\footnotesize
C_{\mathrm{move}}(p) =
\sum_{Q_i\in\mathcal{Q}} \rho_i C_{\mathrm{sep}}(Q_i,p).
\label{eq:movement}
\end{equation}
\sys{} minimizes recurring movement while limiting reorganization:
\figspace\begin{equation}
\footnotesize
\min_{p}\quad C_{\mathrm{move}}(p) + \eta C_{\mathrm{mig}}(p,p_0),
\label{eq:placement}
\end{equation}
subject to per-resource capacity. Here $p_0$ is the current placement,
$C_{\mathrm{mig}}$ is its change cost (\S\ref{sec:migration}), and
$\eta$ trades movement reduction against reorganization during a full
re-solve, such as initial placement or periodic reoptimization. LPMM's
continuous per-group test instead uses an expected query horizon $H$,
estimated from each group's observed rate rather than a fixed weight.
The optimization unit is neither an ANN neighborhood nor an index
structure, but the set of heterogeneous representations that a workload
repeatedly consumes together, at whatever granularity can safely move.
\figspace\figspace

\subsection{Cross-Modal Access Graph}
\label{sec:cag} \figspace

Semantic affinity alone does not establish that two representations
will be consumed together~\cite{zhang2025gme,cui2025dekr,caffagni2025ret,
duan2025fume,elsharief2025medmod}, while sparse observations can be
insufficient for emerging relationships. \sys{} therefore combines
observed co-access with semantic and temporal evidence in a weighted graph
$G_A=(V,E_A,W_A)$, where each vertex is a placement unit. For each pair
$u,v$, CAG maintains co-access $A_{uv}$, semantic affinity $S_{uv}$, and
temporal proximity $T_{uv}$, fused as
\figspace\begin{equation}
w_{uv} = \alpha A_{uv} + \beta S_{uv} + \gamma T_{uv},
\label{eq:cag}
\end{equation}
with tunable $\alpha,\beta,\gamma$. Co-access provides the primary
evidence, semantic affinity provides cold-start evidence, and temporal
proximity captures short-lived correlations. This weighted fusion
provides a practical locality score rather than a probabilistic model.

CAG uses an incremental maintenance policy. Each new access event marks
its object \emph{dirty}, and a rebuild recomputes $A_{uv}$/$S_{uv}$/$T_{uv}$
only for pairs touching a dirty object. Since embeddings are static,
$S_{uv}$ is cached permanently. All other edges remain unchanged: co-access
pairs form only within one object, and $T_{uv}$ depends only on the two
endpoints' own last-access times.

One caveat is that the normalization denominator of $A_{uv}$ is a
\emph{global} statistic, computed from a robust percentile of co-access
counts to reduce sensitivity to individual high-traffic pairs. Consequently,
the statistic can change even when the updated workload does not directly
involve a particular object. When such a change invalidates the existing
normalization, the implementation falls back to a full rebuild to avoid
using stale co-access weights. Thus, the benefit of incremental maintenance
is most pronounced under frequent, small workload updates and progressively
diminishes toward full-rebuild cost for large, infrequent workload shifts.
The \textit{Control-Plane Overhead} evaluation in \S\ref{sec:evaluation}
examines both regimes rather than assuming that the incremental bound holds
unconditionally.\figspace\figspace

\begin{algorithm} [!t]
\caption{CPO: Movement-Aware Group Placement}
\label{alg:cpo}
\begin{algorithmic}[1]
\Require CAG $G_A=(V,E_A,W_A)$, resources $\mathcal{H}$ with capacities,
  vector sizes, current placement $p_0$, optional query-frequency $f(\cdot)$
\Ensure placement $p$
\State $K \gets$ ClusterViaLabelPropagation($G_A$) \Comment{seeded, deterministic}
\If{$f$ given}
  \State sort $K$ by $\bigl(\sum_{v \in K_i} f(v),\ \text{IncidentWeight}(K_i)\bigr)$, descending
\Else
  \State sort $K$ by IncidentWeight($K_i$), descending
\EndIf
\State $Q \gets$ queue initialized with $K$ in sorted order
\While{$Q$ not empty}
  \State $K_i \gets Q$.pop()
  \State $\mathit{size} \gets \sum_{v \in K_i} \mathrm{size}(v)$
  \State $h^\star \gets \bot$; $\mathit{best} \gets (\infty, \infty)$
  \For{$h \in \mathcal{H}$ with remaining capacity $\geq \mathit{size}$}
    \State $\mathit{cost} \gets \sum_{v \in K_i} \sum_{(v,u) \in E_A,\, u \text{ already placed}} w_{vu} \cdot D(h, p(u))$
    \If{$(\mathit{cost}, \text{RefCost}(h)) < \mathit{best}$} \Comment{RefCost = mean $D$ to every other resource}
      \State $h^\star \gets h$; $\mathit{best} \gets (\mathit{cost}, \text{RefCost}(h))$
    \EndIf
  \EndFor
  \If{$h^\star \neq \bot$}
    \State $p(v) \gets h^\star$ for all $v \in K_i$; deduct $\mathit{size}$ from $h^\star$'s remaining capacity
  \ElsIf{$|K_i| \leq 1$} \Comment{no resource had room for even one unit}
    \State $h^\star \gets$ lowest-RefCost resource with any room; else highest-remaining-capacity resource
    \State $p(v) \gets h^\star$ for the sole $v \in K_i$
  \Else
    \State $(v_1,\ldots,v_k) \gets K_i$ sorted by CAG degree, ascending
    \State $Q$.push($\{v_1,\ldots,v_{k-1}\}$); $Q$.push($\{v_k\}$) \Comment{peel off the highest-degree vertex}
  \EndIf
\EndWhile
\State \Return $p$
\end{algorithmic}
\end{algorithm}

\subsection{Movement-Aware Cross-Modal Placement}
\label{sec:cpo} 

Existing graph-partitioning and vector-placement
techniques~\cite{zhi2025cotra,chen2025redanns,fang2026dhnsw,
shim2025turbocharging,ren2025polystore,zhang2020optimal} optimize
partition balance, index locality, or resource utilization, but not the
physical cost of separating representations consumed together by a
multimodal query. \sys{} instead uses cross-modal movement as the
placement objective.

CPO consumes CAG and topology information to determine where placement
units reside. For $v$ at $h_0$, the reduction in future movement cost from
placing it at $h$ is
\begin{equation}
\footnotesize
\Delta C(v,h) =
C_{\mathrm{future}}(v,h_0) - C_{\mathrm{future}}(v,h).
\label{eq:benefit}
\end{equation}
CPO operates on \emph{groups} $K$ of high-connectivity units, so
$\Delta C(K,h)=\sum_{v\in K}\Delta C(v,h)$.

The resulting constrained graph-partitioning problem is NP-hard, so CPO
uses a lightweight heuristic (Protocol~\ref{alg:cpo}). It forms groups
using seeded asynchronous label propagation, orders them by incident
co-locality weight, and assigns each group to a feasible resource that
maximizes $\Delta C(K,h)$ under capacity. Ties are resolved using the
resource's mean cost to other resources rather than incidental
enumeration order.

Incident weight measures connectivity but not query volume. \sys{}
therefore optionally accepts per-vector query frequency and uses aggregate
query volume to order groups when supplied. High-traffic groups claim
fast-tier capacity first, leaving data structurally forced onto slower
tiers disproportionately cold. This ordering does not change
$\Delta C(K,h)$ itself.

When a group cannot fit on one resource, CPO splits it iteratively along
its weakest locality edges and returns the resulting sub-groups to the
queue. CPO never rewrites ANN graph edges, vector codes, or index
structures; it only relocates placement units, preserving each index's
internal optimizations. \figspace\figspace

\subsection{Topology- and Tier-Aware Placement}
\label{sec:topology} \figspace

Generic locality abstractions are insufficient on heterogeneous systems,
where NUMA-remote, CPU--GPU, NVMe, and remote-node costs differ
substantially~\cite{ren2025polystore,zhang2020optimal}. \sys{}
therefore represents the platform as a topology-aware cost matrix in
time units, allowing it to compose with $\rho_i$ in Eq.~\ref{eq:movement}:
\begin{equation}
\footnotesize
D(h_i,h_j) = \lambda_L \hat{L}_{ij}
+ \lambda_B \frac{1}{\hat{BW}_{ij}}
+ \lambda_I \hat{IO}_{ij}
+ \lambda_G \hat{G}_{ij},
\label{eq:topology}
\end{equation}
with normalized latency, bandwidth, I/O, and GPU-transfer terms weighted
by $\lambda_L,\lambda_B,\lambda_I,\lambda_G$. This allows CPO to
distinguish same-NUMA from cross-NUMA access, GPU transfer from NVMe
fetches, and local from remote movement. Thus, the same CAG relationship
can yield different placements under different topologies without
restructuring any ANN index, since only the backing extent moves.
\figspace

\subsection{Locality-Preserving Migration}
\label{sec:migration} \figspace

Static placement~\cite{tang2024tiresias,zhong2024managing,jeong2025lacx}
degrades as access patterns evolve, while continuous reorganization can
outweigh the resulting locality benefits~\cite{zhang2020optimal,jeong2025lacx}.
LPMM monitors CAG changes and evaluates whether units or groups should
move. Migrating $v$ from $h_0$ to $h$ costs
\begin{equation}
\footnotesize
C_{\mathrm{mig}}(v,h) =
\frac{size(v)}{\hat{BW}(h_0,h)},
\label{eq:migration-cost}
\end{equation}
commensurate with $\Delta C(v,h)$. LPMM computes
\begin{equation}
\footnotesize
B(v,h) =
\Delta C_{\mathrm{future}}(v,h) - C_{\mathrm{mig}}(v,h),
\label{eq:migration-benefit}
\end{equation}
and migrates only when $B(v,h)>0$
\label{eq:migration-condition}. For an expected horizon $H$, this is
equivalent to
\begin{equation}
\footnotesize
H\cdot\Delta C_{\mathrm{query}}(v,h) >
C_{\mathrm{mig}}(v,h),
\label{eq:amortization}
\end{equation}
with $\Delta C_{\mathrm{query}}=\Delta C_{\mathrm{future}}/H$ the
expected per-query benefit and $H$ estimated from $v$'s observed rate.
Migration is therefore an amortized decision rather than an unconditional
response to every locality change.

LPMM re-optimizes only units or groups whose relationships changed enough
to matter, avoiding repeated reorganization of cold data. Migration
follows a copy--verify--swap protocol (Protocol~\ref{alg:lpmm}): the
destination is populated and verified, the placement map is updated
atomically, and the old extent is released only after success. A failed
migration therefore leaves the previous placement valid.
\figspace\figspace

\begin{algorithm} [!t]
\caption{LPMM: Threshold-Gated Migration}
\label{alg:lpmm}
\begin{algorithmic}[1]
\Require candidates $C$ (dirty units flagged since the last check), CAG
  $G_A$, current placement $p_0$, $\mathit{margin}$ (default 0)
\Ensure placement $p$, migrated bytes
\State $N \gets C \cup \{u : (v,u) \in E_A,\ v \in C\}$ \Comment{candidates + their CAG neighbors}
\State $G_N \gets$ subgraph of $G_A$ induced on $N$
\State $p' \gets$ CPO($G_N$, \ldots, current placement $= p_0|_N$, pinned $= N \setminus C$) \Comment{Alg.~\ref{alg:cpo}, restricted to $N$}
\State $C_{\mathrm{move}}^{\mathrm{old}} \gets \sum_{(u,v) \in E_{G_N}} w_{uv} \cdot D(p_0(u), p_0(v))$
\State $C_{\mathrm{move}}^{\mathrm{new}} \gets \sum_{(u,v) \in E_{G_N}} w_{uv} \cdot D(p'(u), p'(v))$
\State $\Delta C \gets C_{\mathrm{move}}^{\mathrm{old}} - C_{\mathrm{move}}^{\mathrm{new}}$
\State $C_{\mathrm{mig}} \gets \sum_{v \,:\, p'(v) \neq p_0(v)} \mathrm{size}(v) \,/\, \hat{BW}(p_0(v), p'(v))$
\If{$\Delta C > C_{\mathrm{mig}} + \mathit{margin}$} \Comment{Eq.~\ref{eq:migration-condition}, generalized}
  \For{$v$ with $p'(v) \neq p_0(v)$} \Comment{copy--verify--swap}
    \State copy $v$ to $p'(v)$
    \If{verify($v$ at $p'(v)$) succeeds}
      \State $p(v) \gets p'(v)$ atomically; release $v$'s copy at $p_0(v)$
    \Else
      \State $p(v) \gets p_0(v)$ \Comment{never released -- $p_0$ remains valid for $v$}
    \EndIf
  \EndFor
\Else
  \State $p \gets p_0$ \Comment{decline: below threshold, no migration}
\EndIf
\State \Return $p$
\end{algorithmic}
\end{algorithm}

\subsection{CAG-Driven Prefetching}
\label{sec:prefetch} \figspace

Placement and migration change \emph{where} data resides, but cannot
eliminate the fraction structurally forced onto the slowest tier under
tight capacity (see \textit{end-to-end performance} evaluation in
\S\ref{sec:evaluation}). \sys{} instead hides the cost of
\emph{reaching} that data. When a query accesses $u$, its CAG neighbors
become candidates for background retrieval into a bounded cache,
overlapping the slow tier's round trip with subsequent computation. The
cache is maintained separately from the placement map $p(\cdot)$: a hit
changes where a fetch is served from, but not CPO's or LPMM's record of
where $v$ is placed.

Raw co-access counts do not provide calibrated confidence: two-of-two and
two-hundred-of-two-hundred reflect different levels of evidence that
$A_{uv}$ alone cannot distinguish. \sys{} therefore scores each candidate
$v$ of accessed vertex $u$ under a Beta-Binomial model. Given $s_{uv}$
observed co-accesses and $f_{uv}$ observed accesses of $u$ without $v$
following,
\begin{equation}
P(v \mid u) =
\frac{s_{uv} + a_0}
{s_{uv} + f_{uv} + a_0 + b_0}
\label{eq:beta-binomial}
\end{equation}
under a Beta($a_0,b_0$) prior that shrinks estimates toward the prior for
sparse history and converges to the empirical rate with increasing
evidence. A candidate is prefetched only above an explicit confidence
threshold, and only if $v$ is on a slow tier and not already cached.

Since the worker pool has bounded capacity, dispatch order matters when
more candidates clear the threshold than the pool can serve. \sys{} uses
a confidence-ordered priority queue rather than FIFO submission, so the
pool serves the highest-confidence outstanding candidate first.
\figspace\figspace

\subsection{Operational Overhead and Isolation}
\label{sec:overhead}\figspace 
All optimization mechanisms stay off the query critical path: execution
generates only lightweight access observations, while CAG updates, CPO,
topology evaluation, LPMM migration, and prefetching (\S\ref{sec:prefetch})
run asynchronously in the background -- a prefetch cache miss simply
falls through to the vector's normal placed resource at no extra cost.
The design also avoids global reorganization: CPO evaluates only
affected groups, and LPMM migrates only units whose savings justify
their cost, so \emph{migration decisions scale with the amount of
workload change}, not dataset size.

Incremental CAG maintenance (\S\ref{sec:cag}) is a second, narrower
reduction: it bounds rebuild cost by workload drift only under
\emph{frequent, small} updates, where the global co-access
normalization stays stable between calls. A single large, infrequent
shift -- such as the workload shift in \textit{Adaptive
Migration} evaluation in \S\ref{sec:evaluation} -- can
still trigger a full rebuild. \sys{} therefore bounds control-plane
overhead by workload drift in the frequent-update regime specifically,
not unconditionally. The Evaluation section reports both regimes rather
than assume the bound holds everywhere convenient. \figspace \figspace

\begin{table}[!t]
\centering
\caption{\textbf{Movement-Aware Placement.}}\figspace 
\label{tab:contrib2-placement}
\footnotesize
\begin{tabular}{lrr}
\toprule
\textbf{Metric} & \textbf{Independent} & \textbf{CAG+CPO} \\
\midrule
Cross-tier access rate (\%)   & 73.04  & \textbf{0.00} \\
Bytes on remote tier (\%)     & \textbf{34.9}   & 38.8 \\
End-to-end latency (ms)       & 12.52 & \textbf{11.67} \\
Max tier utilization (\%)     & 166.7  & \textbf{100.0} \\
Capacity constraint violated? & Yes    & \textbf{No} \\
\bottomrule
\end{tabular}
\end{table}

\begin{table}[!t]
\centering
\caption{\textbf{Latency vs.\ Query Intensity.}} \figspace
\label{tab:contrib2-intensity}
\footnotesize
\begin{tabular}{lrrr}
\toprule
\textbf{Multimodal frac.} & \textbf{Indp (ms)} & \textbf{CAG+CPO (ms)} & \textbf{Reduction} \\
\midrule
0.1 & 10.52 & 9.65 & 8.3\% \\
0.3 & 11.64 & 9.73 & 16.4\% \\
0.5 & 13.27 & \textbf{9.98} & \textbf{24.8\%} \\
\bottomrule
\end{tabular} \figspace
\end{table}

\begin{table}[!t]
\centering
\footnotesize

\caption{\textbf{Retrieval Neutrality and End-to-End Cost.}} \figspace
\label{tab:contrib4-ann}
\begin{tabular}{lrrr}
\toprule
\textbf{Configuration} & \textbf{Recall@10} & \textbf{ANN Lat.\ (ms)} & \textbf{E2E Lat.\ (ms)} \\
\midrule
Independent            & 1.0000 & 8.72 & 12.52 \\
Access-Aware            & 1.0000 & 8.72 & 20.67 \\
CAG + CPO               & 1.0000 & 8.72 & 11.67 \\
F. Sys             & 1.0000 & 8.72 & 11.76 \\
F. Sys + Prefetch  & 1.0000 & 8.72 & \textbf{10.75} \\
\bottomrule
\end{tabular}
\end{table}

\begin{table}[!t]
\centering
\caption{\textbf{Adaptive Migration (LPMM).}} \figspace
\label{tab:contrib3-migration}
\footnotesize
\begin{tabular}{lrrr}
\toprule
\textbf{Metric} & \textbf{Static} & \textbf{Periodic Reorg} & \textbf{LPMM} \\
\midrule
Migrated data (MB)              & 0.0   & 30.19 & \textbf{5.10} \\
Migration time (s)               & 0.0   & 54.65 & \textbf{14.03} \\
Post-shift cross-tier rate (\%)  & 42.02 & \textbf{0.00} & 0.11 \\
Post-shift latency (ms)          & 11.16 & 10.82 & \textbf{10.15} \\
\bottomrule
\end{tabular}
\end{table}

\begin{table}[!t]
\centering
\caption{\textbf{Incremental CAG Overhead, Real AWS.}
Many-small-batches regime (Sweep D).} \figspace
\label{tab:contrib4-overhead}
\footnotesize
\begin{tabular}{lr}
\toprule
\textbf{Metric} & \textbf{Value} \\
\midrule
Full-rebuild fallback triggered  & 1 / 19 batches \\
Incremental total time (s)       & 0.756 \\
Full-rebuild-equivalent time (s) & 5.235 \\
Speedup                          & \textbf{6.92$\times$} \\
\bottomrule
\end{tabular} \figspace \figspace
\end{table} 

\section{Evaluation}
\label{sec:evaluation} \figspace 

We implement the prototype in Python for the proposed cross-modal co-locality framework  as a
management layer deployed on AWS, above an existing vector-search stack. The
prototype separates the ANN/search path from the placement-management path,
allowing us to evaluate physical data placement and data movement without
modifying the underlying retrieval procedure. 


\noindent\textbf{Experimental Setup.}
\label{sec:eval-setup} 
We deploy \sys{} on one AWS EC2 instance per tier:
\texttt{g5.xlarge} (GPU), \texttt{r6i.4xlarge} (NUMA-DRAM), \texttt{i4i.2xlarge}
(NVMe), and Amazon S3 (remote, measured 24.2\,ms round-trip, 86.4\,MB/s --
one to two orders of magnitude slower than every other tier). Tier
capacities are sized relative to dataset size (15\%/20\%/30\%/50\% for
GPU/NUMA/NVMe/remote) so no strategy can trivially avoid multi-tier use.
We evaluate 50,000 objects (100,000 vectors, 768 dimensions) from MIMIC-CXR~\cite{johnson2019mimiccxr} (377k radiographs, 228k reports). CAG weights use $\alpha=\beta=\gamma=1.0$
(Eq.~\ref{eq:cag}, uniform weighting). 
We compare five configurations:
\textit{(i) Independent}: modality-specific representations are placed
independently without cross-modal coordination;
\textit{(ii) Access-Aware}: placement uses observed cross-modal access
only, isolating the contribution of workload co-access from semantic and
temporal signals.
\textit{(iii) CAG+CPO}: cross-modal relationships are constructed by CAG
and translated into capacity- and topology-aware placement by CPO.
\textit{(iv) Full System}: CAG+CPO plus LPMM for adaptive migration.
\textit{(v) Full System+Prefetch}: the complete system with
confidence-gated background prefetching.


\noindent\textbf{End-to-End Performance and Movement.}
\label{sec:eval-performance} 
Table~\ref{tab:contrib2-placement} shows that CAG+CPO eliminates cross-tier access (73.04\%$\to$0.00\%) and reduces end-to-end latency by 6.8\% (12.52$\to$11.67,ms), while satisfying the tier-capacity constraints. Although CAG+CPO places slightly more data on the remote tier (38.8\% vs.\ 34.9\%), it achieves lower latency by co-locating frequently accessed representations. In contrast, Independent placement exceeds its capacity allocation (166.7\%), whereas CAG+CPO remains within the limit (100\%). These results demonstrate that movement-aware placement improves cross-modal locality and query performance under constrained heterogeneous storage. Table~\ref{tab:contrib2-intensity} shows that the latency benefit of CAG+CPO increases with multimodal query intensity. The reduction grows from 8.3\% at 10\% multimodal queries to 16.4\% at 30\% and 24.8\% at 50\%, indicating that cross-modal co-locality becomes increasingly beneficial as more queries access multiple modalities.



\noindent\textbf{ANN Quality, Retrieval Neutrality, and Prefetch.}
\label{sec:eval-ann}
Table~\ref{tab:contrib4-ann} shows that Recall@10 (1.0000) and ANN latency (8.72,ms) remain unchanged across all configurations, demonstrating that CoLoc does not alter the ANN retrieval path. Full System+Prefetch further reduces end-to-end latency from 11.76,ms to 10.75,ms (8.5\%) relative to Full System, showing that confidence-gated prefetching can hide remaining remote-access latency without affecting retrieval quality.


\noindent\textbf{Adaptive Migration.}
\label{sec:eval-migration}
Table~\ref{tab:contrib3-migration} shows that LPMM adapts to the workload shift with substantially lower migration cost than periodic reorganization. LPMM moves only 5.10,MB and completes migration in 14.03,s, compared with 30.19,MB and 54.65,s for periodic reorganization---a 5.9$\times$ reduction in data movement and 3.9$\times$ faster migration. At the same time, LPMM nearly restores the locality of periodic reorganization, reducing post-shift cross-tier access from 42.02\% to 0.11\% and achieving the lowest post-shift latency (10.15,ms). These results demonstrate that LPMM effectively adapts placement while substantially reducing the cost of global reorganization.


\noindent\textbf{Control-Plane Overhead.}
\label{sec:eval-overhead}
Table~\ref{tab:contrib4-overhead} shows that incremental CAG maintenance reduces construction time from 5.235,s to 0.756,s, providing a 6.92$\times$ speedup in the many-small-updates regime. The mechanism triggers a full rebuild only once across 19 batches, demonstrating that incremental maintenance substantially reduces control-plane overhead when workload changes are small and frequent.
\figspace \figspace

\section{Conclusion and Future Work}
\label{sec:conclusion} \figspace 
We presented \sys{}, a storage-management layer that treats cross-modal
co-locality as a first-class data-layout objective for distributed vector
stores and comprises three tightly coupled components. CAG captures
cross-modal access, semantic, and temporal locality. CPO translates these
signals into topology- and capacity-aware placement. Finally, LPMM adapts
placement under workload shifts. A confidence-gated prefetcher complements
these components by reducing remaining slow-tier latency. \sys{} operates
above existing ANN indexes without altering retrieval semantics. Our AWS
evaluation demonstrates that cross-modal placement reduces data movement
and end-to-end latency, while LPMM recovers locality after workload shifts
with substantially less migration than global reorganization.

Future work includes online adaptation of placement and prefetching
parameters, evaluation across larger and geographically distributed
deployments, and integration with multi-tenant vector serving.

\bibliographystyle{plain}
\bibliography{ref}

@inproceedings{yin2026distvs,
  author    = {Yin, Zhen and others},
  title     = {DistVS: Large-scale Vector Search},
  booktitle = {Proceedings of the 23rd USENIX Symposium on Networked
               Systems Design and Implementation (NSDI)},
  year      = {2026}
}

@article{gids2025,
  title   = {Design and Optimization of an ANNS Framework Based on GIDS},
  journal = {High-Confidence Computing},
  year    = {2025},
  doi     = {10.1016/j.hcc.2025.100373}
}

@inproceedings{vecflow2025,
  title     = {VecFlow: Accelerating Filtered Approximate Nearest Neighbor
               Search on GPUs},
  year      = {2025},
  booktitle = {Proceedings of the International Conference for High
               Performance Computing, Networking, Storage and Analysis}
}

@inproceedings{yang2025arcade,
  author    = {Yang, Jingyi and Mo, Songsong and Shi, Jiachen and Yu, Zihao
               and Shi, Kunhao and Ding, Xuchen and Cong, Gao},
  title     = {ARCADE: A Real-Time Data System for Hybrid and Continuous
               Query Processing across Diverse Data Modalities},
  journal   = {arXiv preprint arXiv:2509.19757},
  year      = {2025}
}

@article{taipalus2023vectordb,
  title   = {Vector database management systems: Fundamental concepts,
             use-cases, and current challenges},
  author  = {Taipalus, Toni},
  journal = {arXiv preprint arXiv:2309.11322},
  year    = {2023}
}

@article{oldp2024,
  author  = {Zhang, et al.},
  title   = {An Optimized Learning-Based Directory Placement Policy
             with Two-Rounds Selection in Distributed File Systems},
  journal = {Future Generation Computer Systems},
  volume  = {154},
  pages   = {235--250},
  year    = {2024},
  doi     = {10.1016/j.future.2023.12.012}
}

@article{idaps2024,
  author  = {Vengadeswaran, S. and Balasundaram, S. R.
             and Dhavakumar, P.},
  title   = {IDaPS -- Improved Data-Locality Aware Data Placement Strategy
             Based on Markov Clustering to Enhance MapReduce Performance
             on Hadoop},
  journal = {Journal of King Saud University Computer and Information
             Sciences},
  volume  = {36},
  number  = {3},
  pages   = {101973},
  year    = {2024},
  doi     = {10.1016/j.jksuci.2024.101973}
}

@article{adapt2025multicloud,
  author  = {Agyekum, John and Mazumdar, Somnath and Scheich, Christoph},
  title   = {ADAPT: An Effective Data-Aware Multicloud Data Placement
             Framework},
  journal = {Cluster Computing},
  volume  = {28},
  pages   = {825},
  year    = {2025}
}

@misc{bader2025cephs3vectors,
  author       = {Bader, Kyle},
  title        = {Closer Than You Think: Adding Nearest Neighbor Search to Ceph},
  year         = {2025},
  month        = mar,
  howpublished = {Ceph Blog},
  url          = {https://ceph.io/en/news/blog/2026/s3-vectors/}
}

@inproceedings{pinecone2025filtering,
  title     = {Scalable Metadata Filtering for Vector Search},
  year      = {2025},
  booktitle = {Proceedings of the International Conference on Learning
               Representations / OpenReview}
}

@inproceedings{kim2025genius,
  author    = {Kim, Sungyeon and Zhu, Xinliang and Lin, Xiaofan and
               Bastan, Muhammet and Gray, Douglas and Kwak, Suha},
  title     = {GENIUS: A Generative Framework for Universal
               Multimodal Search},
  booktitle = {Proceedings of the IEEE/CVF Conference on Computer Vision
               and Pattern Recognition (CVPR)},
  year      = {2025},
  pages     = {19659--19669}
}

@inproceedings{wang2026mgcs,
  author    = {Wang, Wuyuqing and Gu, Zeyuan and Yang, Erkun},
  title     = {Noisy Correspondence Learning with Modality Gap Direction Correction},
  booktitle = {Proceedings of the AAAI Conference on Artificial Intelligence},
  volume    = {40},
  number    = {12},
  pages     = {10163--10171},
  year      = {2026},
  doi       = {10.1609/aaai.v40i12.37984}
}

@inproceedings{wang2026dns,
  author    = {Wang, Yu and Han, Fengxia and Wang, Jianyu},
  title     = {Boosting Noisy Correspondence Discrimination via Dynamic Neighborhood Semantic Verification},
  booktitle = {Proceedings of the AAAI Conference on Artificial Intelligence},
  volume    = {40},
  number    = {31},
  pages     = {26706--26714},
  year      = {2026},
  doi       = {10.1609/aaai.v40i31.39880}
}

@inproceedings{recon2025,
  author    = {Zha, Quanxing and Liu, Xin and Peng, Shu-Juan and
               Cheung, Yiu-ming and Xu, Xing and Wang, Nannan},
  title     = {ReCon: Enhancing True Correspondence Discrimination
               through Relation Consistency for Robust Noisy
               Correspondence Learning},
  booktitle = {Proceedings of the IEEE/CVF Conference on Computer
               Vision and Pattern Recognition (CVPR)},
  pages     = {29680--29689},
  year      = {2025},
  doi       = {10.1109/CVPR52734.2025.02763}
}

@article{noisycorrespondence2026,
  title   = {Learning with Noisy Cross-Modal Correspondences},
  year    = {2026},
  journal = {IEEE Transactions on Pattern Analysis and Machine Intelligence}
}

@inproceedings{yao2024drfuse,
  title     = {DrFuse: Learning Disentangled Representation for Clinical
               Multi-Modal Fusion with Missing Modality and Modal Inconsistency},
  author    = {Yao, Wenfang and Yin, Kejing and Cheung, William K. and
               Liu, Jia and Qin, Jing},
  booktitle = {Proceedings of the AAAI Conference on Artificial Intelligence},
  volume    = {38},
  number    = {15},
  pages     = {16416--16424},
  year      = {2024},
  doi       = {10.1609/aaai.v38i15.29578}
}

@InProceedings{pmlr-v219-lee23a,
  title     = {Learning Missing Modal Electronic Health Records with Unified Multi-modal Data Embedding and Modality-Aware Attention},
  author    = {Lee, Kwanhyung and Lee, Soojeong and Hahn, Sangchul and
               Hyun, Heejung and Choi, Edward and Ahn, Byungeun and
               Lee, Joohyung},
  booktitle = {Proceedings of the 8th Machine Learning for Healthcare Conference},
  pages     = {423--442},
  year      = {2023},
  volume    = {219},
  series    = {Proceedings of Machine Learning Research},
  publisher = {PMLR}
}

@article{wang2025reliablevdbms,
  author  = {Wang, Shenao and Zhao, Yanjie and Xie, Yinglin and
             Liu, Zhao and Hou, Xinyi and Zou, Quanchen and Wang, Haoyu},
  title   = {Towards Reliable Vector Database Management Systems:
             A Software Testing Roadmap for 2030},
  journal = {arXiv preprint arXiv:2502.20812},
  year    = {2025},
  doi     = {10.48550/arXiv.2502.20812}
}

@article{xie2025vdbbugs,
  author  = {Xie, Yinglin and Hou, Xinyi and Zhao, Yanjie and
             Wang, Shenao and Chen, Kai and Wang, Haoyu},
  title   = {Toward Understanding Bugs in Vector Database
             Management Systems},
  journal = {arXiv preprint arXiv:2506.02617},
  year    = {2025},
  doi     = {10.48550/arXiv.2506.02617}
}

@inproceedings{ren2025polystore,
  author    = {Yujie Ren and David Domingo and Jian Zhang and Paul John
               and Rekha Pitchumani and Sanidhya Kashyap and Sudarsun Kannan},
  title     = {{PolyStore}: Exploiting Combined Capabilities of Heterogeneous Storage},
  booktitle = {23rd USENIX Conference on File and Storage Technologies (FAST 25)},
  year      = {2025},
  pages     = {539--555},
  publisher = {USENIX Association}
}

@article{zhang2020optimal,
  author  = {Lei Zhang and Reza Karimi and Irfan Ahmad and Ymir Vigfusson},
  title   = {Optimal Data Placement for Heterogeneous Cache, Memory, and Storage Systems},
  journal = {Proceedings of the ACM on Measurement and Analysis of Computing Systems},
  volume  = {4},
  number  = {1},
  pages   = {1--27},
  year    = {2020},
  doi     = {10.1145/3379472}
}

@inproceedings{johnson2019mimiccxr,
  author    = {Alistair E. W. Johnson and Tom J. Pollard and Seth J. Berkowitz and Nathaniel R. Greenbaum and Matthew P. Lungren and Chih-ying Deng and Roger G. Mark and Steven Horng},
  title     = {{MIMIC-CXR}, a de-identified publicly available database of chest radiographs with free-text reports},
  journal   = {Scientific Data},
  volume    = {6},
  pages     = {317},
  year      = {2019}
}

@inproceedings{zhi2025cotra,
  author    = {Xiangyu Zhi and Meng Chen and Xiao Yan and Baotong Lu and Hui Li and Qianxi Zhang and Qi Chen and James Cheng},
  title     = {Towards Efficient and Scalable Distributed Vector Search with {RDMA}},
  booktitle = {Proceedings of the ACM on Management of Data (SIGMOD/PACMMOD)},
  volume    = {4},
  year      = {2025}
}

@article{chen2025redanns,
  author  = {Yue Chen and Kai Zhang and Sipeng Chen and Shihai Xiao and Xiaomin Zou and Ren Ren and Yinan Jing and X. Sean Wang and Li Cao and Mingxiang Wan},
  title   = {{RED-ANNS}: An {RDMA}-Enabled Distributed Framework for Graph-Based Approximate Nearest Neighbor Search},
  journal = {Proceedings of the VLDB Endowment},
  volume  = {19},
  number  = {3},
  pages   = {399--412},
  year    = {2025}
}

@article{fang2026dhnsw,
  author  = {Fei Fang and Yi Liu and Chen Qian},
  title   = {d-{HNSW}: A High-performance Vector Search Engine on Disaggregated Memory},
  journal = {arXiv preprint arXiv:2603.13591},
  year    = {2026}
}

@article{suri2026vextra,
  author  = {Chandan Suri and Gursifath Bhasin},
  title   = {Vextra: A Unified Middleware Abstraction for Heterogeneous Vector Database Systems},
  journal = {arXiv preprint arXiv:2601.06727},
  year    = {2026}
}

@inproceedings{zhong2024managing,
  author    = {Yuhong Zhong and Daniel S. Berger and Carl A. Waldspurger and Ryan Wee and Ishwar Agarwal and Rajat Agarwal and Frank Hady and Karthik Kumar and Mark D. Hill and Mosharaf Chowdhury and Asaf Cidon},
  title     = {Managing Memory Tiers with {CXL} in Virtualized Environments},
  booktitle = {18th USENIX Symposium on Operating Systems Design and Implementation (OSDI 24)},
  pages     = {37--56},
  year      = {2024}
}

@inproceedings{tang2024tiresias,
  author    = {Wenda Tang and Tianxiang Ai and Jie Wu},
  title     = {Tiresias: Optimizing {NUMA} Performance with {CXL} Memory and Locality-Aware Process Scheduling},
  booktitle = {ACM Turing Award Celebration Conference (ACM-TURC)},
  year      = {2024}
}

@article{jeong2025lacx,
  author  = {H. Jeong and B. Song and M. Jo and H. Jo},
  title   = {{LACX}: Locality-Aware Shared Data Migration in {NUMA} + {CXL} Tiered Memory},
  journal = {Electronics},
  volume  = {14},
  number  = {21},
  pages   = {4235},
  year    = {2025}
}

@article{song2026vectorsearch,
  author  = {Song, Yitong and Zhou, Xuanhe and Jensen, Christian S. and
             Xu, Jianliang},
  title   = {Vector Search for the Future: From Memory-Resident, Static
             Heterogeneous Storage, to Cloud-Native Architectures},
  journal = {arXiv preprint arXiv:2601.01937},
  year    = {2026}
}

@article{shim2025turbocharging,
  author    = {Shim, Joobo and Oh, Jaewon and Roh, Hongchan
               and Do, Jaeyoung and Lee, Sang-Won},
  title     = {Turbocharging Vector Databases Using Modern SSDs},
  journal   = {Proceedings of the VLDB Endowment},
  volume    = {18},
  number    = {11},
  pages     = {4710--4722},
  year      = {2025},
  doi       = {10.14778/3749646.3749724}
}

@inproceedings{zhang2025gme,
  author    = {Zhang, Xin and Zhang, Yanzhao and Xie, Wen and
               Li, Mingxin and Dai, Ziqi and Long, Dingkun and
               Xie, Pengjun and Zhang, Meishan and Li, Wenjie and
               Zhang, Min},
  title     = {Bridging Modalities: Improving Universal Multimodal
               Retrieval by Multimodal Large Language Models},
  booktitle = {Proceedings of the IEEE/CVF Conference on Computer Vision
               and Pattern Recognition (CVPR)},
  year      = {2025},
  pages     = {9274--9285}
}

@inproceedings{cui2025dekr,
  author    = {Cui, Yuhao and Zu, Xinxing and Zhang, Wenhua
               and Zhao, Zhongzhou and Gao, Jinyang},
  title     = {Incorporating Dense Knowledge Alignment into Unified
               Multimodal Representation Models},
  booktitle = {Proceedings of the IEEE/CVF Conference on Computer Vision
               and Pattern Recognition (CVPR)},
  pages     = {29733--29743},
  year      = {2025}
}

@inproceedings{caffagni2025ret,
  author    = {Caffagni, Davide and Sarto, Sara and Cornia, Marcella
               and Baraldi, Lorenzo and Cucchiara, Rita},
  title     = {Recurrence-Enhanced Vision-and-Language Transformers
               for Robust Multimodal Document Retrieval},
  booktitle = {Proceedings of the IEEE/CVF Conference on Computer Vision
               and Pattern Recognition (CVPR)},
  pages     = {9286--9295},
  year      = {2025}
}

@inproceedings{duan2025fume,
  author    = {Duan, Siyuan and Sun, Yuan and Peng, Dezhong and Liu, Zheng
               and Song, Xiaomin and Hu, Peng},
  title     = {Fuzzy Multimodal Learning for Trusted Cross-modal Retrieval},
  booktitle = {Proceedings of the IEEE/CVF Conference on Computer Vision
               and Pattern Recognition (CVPR)},
  pages     = {20747--20756},
  year      = {2025}
}

@article{pan2024vectordb,
  author  = {Pan, James Jie and Wang, Jianguo and Li, Guoliang},
  title   = {Survey of Vector Database Management Systems},
  journal = {The VLDB Journal},
  volume  = {33},
  pages   = {1591--1615},
  year    = {2024},
  doi     = {10.1007/s00778-024-00864-x}
}

@article{lu2025mmstorage,
  author  = {Lu, Yihao and Tang, Hao},
  title   = {Multimodal Data Storage and Retrieval for Embodied AI:
             A Survey},
  journal = {arXiv preprint arXiv:2508.13901},
  year    = {2025},
  doi     = {10.48550/arXiv.2508.13901}
}

@inproceedings{elsharief2025medmod,
  title     = {MedMod: Multimodal Benchmark for Medical Prediction Tasks
               with Electronic Health Records and Chest X-Ray Scans},
  author    = {Elsharief, Shaza and Shurrab, Saeed and Jorf, Baraa Al and
               Lopez, Leopoldo Julian Lechuga and Geras, Krzysztof J. and
               Shamout, Farah E.},
  booktitle = {Proceedings of the sixth Conference on Health, Inference,
               and Learning},
  volume    = {287},
  pages     = {781--803},
  year      = {2025}
}
\end{document}